\documentclass[showpacs,showkeys]{revtex4}
\usepackage{amssymb}
\usepackage{amsmath}
\usepackage{amsfonts}
\usepackage{graphicx}
\providecommand{\rbr}[1]{\left( #1 \right)}%
\providecommand{\sqbr}[1]{\left[ #1 \right]} %
\providecommand{\mt}[1]{\mathrm{#1}}%
\providecommand{\mc}[1]{\mathcal{#1}}%
\providecommand{\tprod}[1]{\sideset{}{_{\otimes_{#1}}}\prod}%

\def\ra{\rightarrow}
\def\BG{S_{\mathrm{BGS}}}
\def\Ts{S_q^{\mathrm{T}}}
\def\Re{S_r^{\mathrm{R}}}
\def\NeG{S_q^{\mathrm{G}}}

\begin{document}

\title[ ]{Generalized entropic structures and non-generality of Jaynes' Formalism}
\author{Thomas Oikonomou}
\email{thoikonomou@chem.demokritos.gr}
\affiliation{Institute of Physical Chemistry, National Center for Scientific Research ``Demokritos", 15310 Athens, Greece}

\author{Ugur Tirnakli}
\email{ugur.tirnakli@ege.edu.tr}
\affiliation{Department of Physics, Faculty of Science, Ege University, 35100 Izmir, Turkey}

\begin{abstract}
The extremization of an appropriate entropic functional may yield to the probability distribution functions maximizing the respective entropic structure.
This procedure is known in Statistical Mechanics and Information Theory as Jaynes' Formalism and has been up to now a standard methodology for deriving the aforementioned distributions.
However, the results of this formalism do not always coincide with the ones obtained following different approaches.
In this study we analyse these inconsistencies in detail and demonstrate that Jaynes' formalism leads to correct results only for specific entropy definitions.
\end{abstract}
\eid{ }
\date{\today }
\startpage{1}
\endpage{1}
\maketitle

\section{Introduction}\label{intro}
%
Jaynes' Formalism (JF) \cite{Jaynes1957} of the maximum entropy principle applied on the Boltzmann-Gibbs-Shannon (BGS)
entropy \cite{Shannon}
%
\begin{align}\label{BGS}
S_{\mt{BGS}}(p_i)&=\ln\rbr{\prod_{i=1}^{\Omega_{\mt{max}}^{\mt{BGS}}}(1/p_i)^{p_i}}
=\sum_{i=1}^{\Omega_{\mt{max}}^{\mt{BGS}}}p_i\ln(1/p_i)
\end{align}
%
reproduce correctly the exponential maximum Probability Distribution Functions (PDFs), obtained from the theory of
thermodynamics.
$\Omega_{\mt{max}}^{\mt{BGS}}$ is the maximum configuration function of the BGS-statistical ensemble and $p_i$ are the associated configuration probabilities.
This celebrated result has established the above formalism as a standard procedure for the computation of the maximum PDFs for an arbitrary generalized entropic structure $S_{\mc{Q}}$ \cite{PlastinoPlastino1997,Wang02,Parvan05,JizbaArim06,FrankPlastino02}, where $\mc{Q}=\{Q_i\}_{i=1,\ldots,n}$
is a set of parameters.
For specific values of these parameters, $\mc{Q}\ra\mc{Q}_0$, the quantity $S_{\mc{Q}}$ tends to $\BG$ in Eq. (\ref{BGS}).
In JF the probability functional
%
\begin{align}\label{JF1}
\mc{I}(p_i)&=S_{\mc{Q}}(p_i)-\alpha\sum_{i}p_i-\beta\sum_{i}E_i\,g(p_i)
\end{align}
%
is extremized, namely, $\delta\mc{I}(p_i)=0$.
The constants $\alpha$ and $\beta$ are the Lagrange multipliers associated to the normalization and mean value constraint ($\sum_{i}p_i$, $\sum_{i}E_i\,g(p_i)$), respectively.
For $g(p_i):=p_i$ we have the Ordinary Mean Value (OMV) definition.
In the frame of generalized thermostatistics \cite{Tsallis88,ts2,ts3} a common definition of the mean value is $g(p_i):=p_i^q/\sum_kp_k^q$, where $q$ is the generalization parameter of the respective entropy.
This definition is called Escort Mean Value (EMV).
$E_{i}$ is the observed quantity for a system under consideration.

Some  well known generalized entropic structures, which have been explored within JF are the one-parametric Tsallis $\Ts$ \cite{Tsallis88}, R\'enyi $\Re$ \cite{Renyi1970} and Nonextensive Gaussian (NeG) $\NeG$ \cite{FrankDaff2000}
ones, defined as
%
\begin{align}
\label{TsallisEntropy}%
\Ts&=\sum_{i=1}^{\Omega_{\mt{max}}^{\mt{T}}} p_i\ln^{\mt{T}}_q(1/p_i)
\qquad\quad\,\,\;\rbr{S_{q\ra1}^{\mt{T}}=\BG},\\
\label{RenyiEntropy}%
\Re&=\frac{\ln\rbr{\sum_{i=1}^{\Omega_{\mt{max}}^{\mt{R}}} p_i^r}}{1-r}
\qquad\quad\quad\rbr{S_{r\ra1}^{\mt{R}}=\BG},\\
\label{NeGEntropy}%
\NeG &=\ln^{\mt{T}}_q\rbr{\prod_{i=1}^{\Omega_{\mt{max}}^{\mt{G}}}(1/p_i)^{p_i}}
\qquad\rbr{S_{q\ra1}^{\mt{G}}=\BG},
\end{align}
%
%
where $\Omega_{\mt{max}}^{\mt{T,R,G}}$ are the respective maximum configuration functions  and $\ln^{\mt{T}}_q(x)$ is the deformed logarithm defined by Tsallis and collaborators
%
\begin{align}\label{TsallisLogarithmicFunction}
\ln^{\mt{T}}_q(x)&:=\frac{x^{1-q}-1}{1-q},\qquad \rbr{x>0,\,\ln^{\mt{T}}_{q\ra1}(x)=\ln(x)}.
\end{align}
%
All of these entropies are based on the generalized logarithmic function (\ref{TsallisLogarithmicFunction}) and its
inverse function
%
\begin{align}
\label{TsallisExponentialFunction1}%
\exp^{\mt{T}}_q(x)&:=\sqbr{1+(1-q)x}_{+}^{\frac{1}{1-q}}\qquad \rbr{\lim_{q\ra1}\exp^{\mt{T}}_{q}(x)=\exp(x)},
\end{align}
with $\sqbr{X}_+=\mt{max}\{0,X\}$.
Indeed, considering the deformed $q$-product \cite{Borges04,wang,Oik2007a} defined within generalized thermostatistics, we find the mathematical relation
%
\begin{align}
\label{q-stat}
\tprod{r}_i(1/p_i)^{\otimes_r^{p_i}}&=\exp^{\mt{T}}_r\rbr{\sum_ip_i\ln^{\mt{T}}_r(1/p_i)}
                                     =\sqbr{\sum_ip_i^r}^{\frac{1}{1-r}}.
\end{align}
%
Comparing Eq.~(\ref{q-stat}) with Eq.~(\ref{BGS}) we see that the earlier is the $r$-generalized inner structure of
the BGS-entropy.
The application of the deformed logarithm (\ref{TsallisLogarithmicFunction}) on Eq.~(\ref{q-stat}) leads to R\'enyi, Tsallis and NeG-entropy for $q\ra1$, $r\ra q$ and $r\ra1$ respectively.

The application of Eq.~(\ref{JF1}) on $\Ts$ and $\Re$ using OMV and EMV yields a $q$-exponential as in Eq.~(\ref{TsallisExponentialFunction1}) and a ($2-q$)-exponential PDF \cite{TsallisMendesPlastino,LenziMendesSilva}, respectively
%
\begin{align}
\label{TsallisExponentialFunction2}%
\exp^{\mt{T}}_{2-q}(x)&:=\sqbr{1+(q-1)x}_{+}^{\frac{1}{q-1}}\qquad \rbr{\lim_{q\ra1}\exp^{\mt{T}}_{2-q}(x)=\exp(x)}.
\end{align}
%
This result may be verified through the monotonic relation between $\Ts$ and $\Re$ for $r=q$ and
$\Omega_{\mt{max}}^{\mt{T}}=\Omega_{\mt{max}}^{\mt{R}}$
%
%
\begin{align}\label{RenyiTsallis}
S_{q}^{\mt{R}}&=\ln\sqbr{\exp_q\rbr{\Ts}},
\end{align}
%
which is often used in literature \cite{TsallisBqrigatti,Gorban,BatlePCP}.
In case of $\NeG$, one obtains from JF using Eq.~(\ref{JF1}) with EMV, a PDF of the form $q_{\mc{L}}$-exponential \cite{Oik-NeG}
%
\begin{align}\label{LambertExp}
\exp^{\mc{L}}_{q}(x)&:=\exp\rbr{\frac{W[(1-q)x]_+}{1-q}} \qquad \rbr{\exp^{\mc{L}}_{q\ra1}(x)=\exp(x)},
\end{align}
%
where $W[x]$ is the $W$-Lambert function \cite{Corless1996}.
The NeG-entropy has not been studied explicitly for OMV.
However, it is easy to see that in this case one would obtain ordinary exponential distributions, since
%
\begin{align}\label{LambertExp}
\frac{d\NeG(p_i)}{dp_i}&\sim\sum_i(1+\ln(p_i)).
\end{align}
%

In a recent study \cite{Oik2007a} one of us (TO) has constructed two Generalized Multinomial Coefficients (GMC),
$\mt{C}_{\{\odot,\mc{R},\mc{Q}\}}^{N,p_i}$ and $\mt{C}_{\{\otimes,\mc{R},\mc{Q}\}}^{N,p_i}$, based on different
generalized factorial operators, $x!_{\{\odot_{\mc{R}}\}}$ and $x!_{\{\otimes_{\mc{R}}\}}$, from which $\Ts$, $\Re$ and $\NeG$ may be derived.
$\mc{Q}$ and $\mc{R}$ are two sets of parameters, $\mc{Q}=\{Q_i\}_{i=1,\ldots,n}$ and $\mc{R}=\{R_i\}_{i=1,\ldots,m}$.
Then, the relation between these parameter sets and the  parameters in $\Ts$, $\NeG$ and $\Re$ is given by $\mc{Q}=Q_1=q$ and $\mc{R}=R_1=r$.
The results about the maximum PDFs for the Tsallis entropy computed from $\mt{C}_{\{\odot,\mc{R},\mc{Q}\}}^{N,p_i}$ are in accordance with Eq.~(\ref{TsallisExponentialFunction1}) and computed from $\mt{C}_{\{\otimes,\mc{R},\mc{Q}\}}^{N,p_i}$ are in accordance with Eq.~(\ref{TsallisExponentialFunction2}) \cite{Oik2007a}.
Furthermore, in both cases the $q$-ranges were determined, namely $q\in[0,1]$ and $q\geqslant1$, respectively.
However, in the case of R\'enyi and NeG entropy the results obtained from GMC are different from the ones obtained from JF.
The R\'enyi entropy is maximized for ordinary exponential distributions (with $r\in[0,1]$ for both GMC's) and the
Nonextensive Gaussian entropy is maximized for $q$- and $(2-q)$-exponential distributions in Eqs. (\ref{TsallisExponentialFunction1}) and (\ref{TsallisExponentialFunction2}), for $q\leqslant1$ and $q\geqslant1$, respectively \cite{Oik2007a}.
These GMC results have a consequence that Eq.~(\ref{RenyiTsallis}) is physically ($r\ne q$) and mathematically
($\Omega_{\mt{max}}^{\mt{T}}\ne\Omega_{\mt{max}}^{\mt{R}}$) incorrect.

The above mentioned discrepancies in the last paragraph indicate a problem in the mathematical structure either
of JF or of GMC, since they may yield different results computing the maximum PDF of an entropy definition.
In the present manuscript we would like to explore which mathematical approach gives proper results and determine
the origin of the problem in the respective approach.
In Section \ref{Sec:2}, based on the concept of extensivity, we argue why the results of JF are in general incorrect.
In Section \ref{sec:3} we investigate the entropic structures on which JF is applicable.
In the last section we draw our conclusions.

\section{GMC, JF and the $\ln_q^{\mt{T}}$-entropies}\label{Sec:2}
%
A general expression of a deformed entropic structure may be given in the following way
%
\begin{align}\label{GenEntStructure}
S_{\mc{Q}}&:=\sum_{i=1}^{\Omega_{\mt{max}}^{(\mc{Q})}}\Lambda_{\mc{Q}}(p_i)
=\sum_{i=1}^{\Omega_{\mt{max}}^{(\mc{Q})}}p_i\frac{\Lambda_{\mc{Q}}(p_i)}{p_i}
=\sum_{i=1}^{\Omega_{\mt{max}}^{(\mc{Q})}}p_i\ln_{\mc{Q}}(1/p_i),
\end{align}
%
where $\ln_{\mc{Q}}(1/p_i):=\Lambda_{\mc{Q}}(p_i)/p_i$ is  a deformed logarithm and for specific values of its parameters, $\mc{Q}\ra\mc{Q}_0$, it tends to the ordinary definition of the logarithmic function.
In shake of simplicity, in what follows we shall consider $S_{\mc{Q}}$ for equal configuration probabilities,
$p_i=1/\Omega_{\mt{max}}^{(\mc{Q})}$, without losing the generality of the results.
Then, we obtain
%
\begin{align}\label{GenEnt1}
S_{\mc{Q}}(\mt{max})&=\ln_{\mc{Q}}(\Omega_{\mt{max}}^{(\mc{Q})}).
\end{align}
%
We recall at this point that a very important property of an entropic form is its extensivity with respect to the
variable under consideration.
For instance, such a variable in thermodynamics may be the size $N$ of a statistical system.
Then, the entropy production of the system must be proportional to $N$.
Taking into account Eq.~(\ref{GenEnt1}), it becomes evident that the maximum configuration function
$\Omega_{\mt{max}}^{(\mc{Q})}$ has to present the inverse structure of the deformed logarithm $\ln_{\mc{Q}}$,
namely, $\exp_{\mc{Q}}$.
If the system under consideration comprises $W$ different types of elements, then $\Omega_{\mt{max}}^{(\mc{Q})}$ takes the form
%
\begin{align}\label{maxConfFun1}
\Omega_{\mt{max}}^{(\mc{Q})}&:=\exp_{\mc{Q}}\rbr{N\phi(\mc{Q})\ln_{\mc{Q}}(W)},
\end{align}
%
under the constraints
%
\begin{align}\label{Constraints1}
\lim_{\mc{Q}\ra\mc{Q}_0}\phi(\mc{Q})&=1,\qquad\lim_{\mc{Q}\ra\mc{Q}_0}\Omega_{\mt{max}}^{(\mc{Q})}=
\exp\rbr{N\ln(W)}=W^N=\Omega_{\mt{max}}^{\mt{BGS}}.
\end{align}
%
Using the GMC approach, Eqs.~(\ref{maxConfFun1}) and (\ref{Constraints1}) have been analytically derived in the
specific case of Tsallis entropy \cite{Oik2007a} and can be shown to be valid for any entropic structure of
the form (\ref{GenEntStructure}) \cite{Oik-progress}.
Replacing Eq.~(\ref{maxConfFun1}) in Eq.~(\ref{GenEnt1}) we obtain the desired extensivity with respect to $N$:
%
\begin{align}\label{maxEntFun1}
S_{\mc{Q}}(\mt{max})&=N\phi(\mc{Q})\ln_{\mc{Q}}(W).
\end{align}
%
As can be seen in Eq.~(\ref{maxEntFun1}), the function $\phi(\mc{Q})$ in thermodynamics may be related to a generalized Boltzmann constant.
Returning to the results of GMC and JF for $\Ts$, $\Re$ and $\NeG$ we make in the following subsections some crucial comments, which have not be considered explicitly in literature up to now.

\subsection{GMC, JF and Tsallis entropy}
%
As discussed in the introduction the application of JF on Tsallis entropy leads to $q$-exponential and ($2-q$)-exponential distributions for ordinary and escort mean values, respectively.
However, it is easy to verify considering $\Ts(\mt{max})$, that PDFs of $q$-exponential type makes $\Ts$ extensive,
since it is the inverse $\ln_q^{\mt{T}}$-function, while PDFs of ($2-q$)-exponential form do not.
According to GMC approach, the $q$-exponential distributions are obtained from the definition $\mt{C}_{\{\odot,r=q,q\}}^{N,p_i}$ with $q\in[0,1]$ and the ($2-q$)-exponential distributions are obtained from the definition $\mt{C}_{\{\otimes,r=q,q\}}^{N,p_i}$ with $q\geqslant1$.
Comparing the results of both approaches, we conclude that the PDFs which maximize $\Ts$ are the $q$-exponential ones,
($p_i=1/\exp_q(x)$ and not $p_i=\exp_q(-x)=1/\exp_{2-q}(x)$), with ordinary mean values and $q\in[0,1]$.
We notice that, the distribution $p_i=\exp_q(-x)=1/\exp_{2-q}(x)$ maximizes the entropy $S_{2-q}^{\mt{T}}$ with ordinary mean values.
>From GMC we determine the range of $q$-values for $S_{2-q}^{\mt{T}}$ which vary between $q\in[1,2]$.
As can be seen, the distinction between $\exp_q$ and $\exp_{2-q}$ is very important since they are related to different entropic structures.
It is a very common mistake in literature to consider a $\exp_q(-x)$ distribution calling it $q$-exponential, although it is a ($2-q$)-exponential distribution.
This is the reason of obtaining in these cases a $q$-value greater than unity.

\subsection{GMC, JF and R\'enyi entropy}
\begin{figure}
\begin{center}
  \includegraphics
  {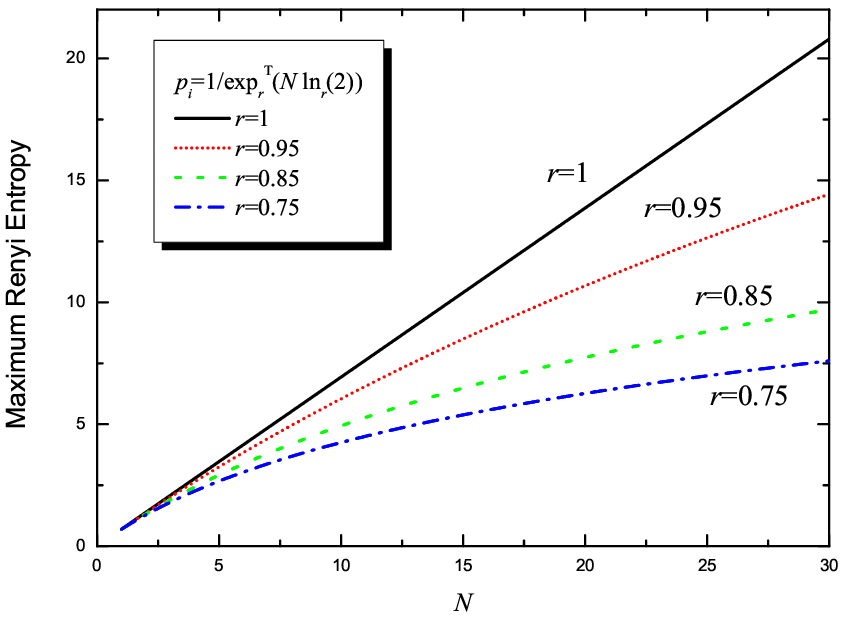}
  \caption{Plots of the entropy $\Re$ for equal configuration probabilities $p^{\mt{JF}}_i=1/\exp_{r}(N\ln_r{(2)})$ and
  $p_i^{\mt{GMC}}=1/\exp_{r\ra1}(N\ln_{r\ra1}{(2)})=2^{-N}$.}%
  \label{fig.Renyi}
\end{center}
\end{figure}

%
Same as in case of $\Ts$, the application of JF on R\'enyi entropy leads to $q$-exponential and ($2-q$)-exponential distributions for ordinary and escort mean values, respectively.
The external function of $\Re(\mt{max})$ is an ordinary logarithm, thus none of the aforementioned distributions makes R\'enyi entropy extensive.
According to GMC approach, for both GMC-definitions $\Re$ is defined for $r\in[0,1]$ and is maximized for ordinary exponential distributions.
These results are in agreement with the concept of extensivity.
In Fig.~\ref{fig.Renyi} we present some plots of the R\'enyi entropy for equal probabilities and $W=2$ types of elements, considering the JF maximum PDFs given by $p_i=1/\exp_{r}(N\ln_r{(2)})$ and the GMC maximum PDFs given by
$p_i=1/\exp_{r\ra1}(N\ln_{r\ra1}{(2)})=2^{-N}$.
As can be seen, and argued above, we obtain extensivity only for $r=1$.
Additionally, when the entropic parameter varies in the range $r\in[0,1)$, the values of $\Re(\mt{max})$ are smaller than the ones for $r=1$.
Thus, PDFs obtained from JF do clearly not maximize the R\'enyi entropy.
%

\subsection{GMC, JF and NeG entropy}
\begin{figure}
\begin{center}
  \includegraphics
  {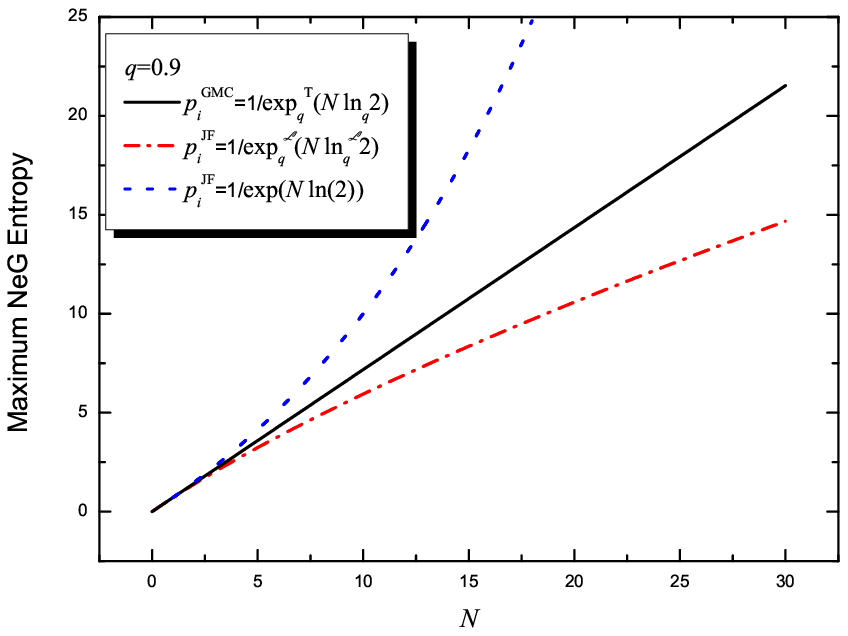}
  \caption{Plots of $\NeG$ for equal probabilities $p^{\mt{GMC}}_i=1/\exp^{\mt{T}}_{q}(N\ln_q{(2)})$,
    $p^{\mt{JF},\mt{EMV}}_i=1/\exp^{\mc{L}}_{q}(N\ln^{\mc{L}}_q{(2)})$ and
    $p^{\mt{JF},\mt{OMV}}_{i}=1/\exp(N\ln{(2)})$ for $q=0.9$.}%
  \label{fig.NeG}
\end{center}
\end{figure}

%
In case of Nonextensive Gaussian entropy one obtains from JF with ordinary and escort mean value constraints ordinary and $q_{\mc{L}}$-exponential distributions.
However, considering $\NeG(\mt{max})=\ln_q^{\mt{T}}(\Omega_{\mt{max}}^{\mt{G}})$, we verify that both distributions do not preserve extensivity.
The PDFs which maximize $\NeG$ are $q$-exponential distributions, as in Eq.~(\ref{TsallisExponentialFunction1}).
Taking into account the results obtained from the GMC, we see that PDFs of $q$-exponential type are derived from the definition $\mt{C}_{\{\odot,r=1,q\}}^{N,p_i}$ with $q\leqslant1$.
In Fig.~\ref{fig.NeG} we present in analogy to Fig.~\ref{fig.Renyi} three plots of the equiprobabilized NeG-entropy,
based on an ordinary, $q$- and $q_{\mc{L}}$-exponential distribution.
Again we see that the extensivity is preserved only for the $q$-exponential PDF, while the values of $\NeG$ based on
the $q_{\mc{L}}$-exponential PDF are lower than the ones based on the $\mt{C}_{\{\odot,r=1,q\}}^{N,p_i}$-PDF, for the
same $q$-value.
Accordingly, the NeG-entropy is optimized by PDFs of $q$-exponential type.

The above discussion implies, with regard to GMC approach, that only the definition $\mt{C}_{\{\odot,r=1,q\}}^{N,p_i}$ gives proper results, and with regard to JF, that the correctness of the obtained maximum PDFs for an entropic structure is not guaranteed.
In the next section we shall demonstrate the origin of the observed inconsistencies within JF.
The computation of the maximum PDF for a given entropic definition and the determination of the parameter range through the respective GMC is not part of the scope of this article and will be explored separately.

\section{JF and generalized logarithms}\label{sec:3}
%
Let us first consider the BGS-entropy in Eq.~(\ref{BGS}).
The inner structure of $\BG$ is of the form $\Lambda(x)=\Lambda_{\mc{Q}_0}(x)=x\ln(1/x)$.
Then, the derivation with respect to the variable $x$ gives
%
\begin{align}\label{derivationBG}
\frac{d}{dx}\Lambda(x)&=\ln(1/x)+x\frac{d}{dx}\ln(1/x)=\ln(1/x)-1.
\end{align}
%
As can be seen in the right hand side of Eq.~(\ref{derivationBG}), the reason of obtaining exponential PDFs in JF for the BGS-entropy, is the remaining logarithmic dependence on $x$, $\ln(1/x)$.
Since the exponential function is the inverse logarithmic function, the extensivity of $\BG$ is preserved
(see Eqs.~(\ref{maxConfFun1}) and (\ref{maxEntFun1}) for $\mc{Q}\ra\mc{Q}_0$).
This observation is of major importance for our further investigation.
It means that JF may lead to correct results if the following two conditions are satisfied:
i)~the entropy definition is of trace-form, as in Eq.~(\ref{GenEntStructure}) and
ii)~the derivation of its inner structure presents the form
%
\begin{align}\label{derivGenFun}
\frac{d}{dx}\Lambda_{\mc{Q}}(x)& =\ln_{\mc{Q}}(1/x)+x\frac{d}{dx}\ln_{\mc{Q}}(1/x) =
f_{1}(\mc{Q})\ln_{\mc{Q}}(1/x)-f_{2}(\mc{Q})
\end{align}
%
with the constraints $\lim_{\mc{Q}\ra\mc{Q}_0}f_{1}(\mc{Q})=\lim_{\mc{Q}\ra\mc{Q}_0}f_{2}(\mc{Q})=1$.
We notice that the  nonsingular functions $f_1(\mc{Q})$ and $f_2(\mc{Q})$ are \emph{independent} from the variable $x$.
The second condition guaranties the extensivity of the entropy $S_{\mc{Q}}$.
>From the middle and right hand side of Eq.~(\ref{derivGenFun}) we can determine the structure of the deformed logarithm $\ln_{\mc{Q}}$.
Substituting $y_{\mc{Q}}(x)=\ln_{\mc{Q}}(1/x)$, $a_{\mc{Q}}(x)=(1-f_1(\mc{Q}))/x$ and $b_{\mc{Q}}(x)=-f_2(\mc{Q})/x$, we obtain
%
\begin{align}
y_{\mc{Q}}'(x)+a_{\mc{Q}}(x)y_{\mc{Q}}(x)&=b_{\mc{Q}}(x).
\end{align}
%
This is a first order ordinary linear differential equation.
Its solution is given as follows
%
\begin{align}\label{Solution1}
y_{\mc{Q}}(x)&=x^{f_1(\mc{Q})-1}c_1-\frac{f_2(\mc{Q})}{1-f_1(\mc{Q})},
\end{align}
%
where $c_1$ is the integration constant.
In the limit $\mc{Q}\ra\mc{Q}_0$ Eq. (\ref{Solution1}) should tend to $y_{\mc{Q}_0}(x)=\ln(1/x)$.
Taking this constraint into account, we can easily verify that the constants $c_1$ and $f_2(\mc{Q})$ must be of the form
%
\begin{align}\label{Approx}
c_1&=\frac{1}{1-f_1(\mc{Q})}\qquad\mt{and}\qquad f_2(\mc{Q})=1.
\end{align}
%
Resubstituting $\ln_{\mc{Q}}(x)\ra y_{\mc{Q}}(1/x)$, we obtain
%
\begin{align}\label{GenLog}
\ln_{\mc{Q}}(x)&=\frac{x^{1-f_1(\mc{Q})}-1}{1-f_1(\mc{Q})}.
\end{align}
%
In other words, JF leads to correct results only when the trace-form entropic structure $S_{\mc{Q}}$ is based on the
deformed logarithm (\ref{GenLog}).
For $\mc{Q}=Q_1=f_1(\mc{Q})=q$ the respective entropy is $\Ts$ in Eq.~(\ref{TsallisEntropy}).
Here, it becomes evident why the results of the JF and GMC approaches coincide in the case of Tsallis entropy,
while for $\Re$ and $\NeG$ they do not.
In further, we shall give some more examples where JF does not gives proper results.

\subsection{Kaniadakis $\kappa$-Entropy}
%
An entropic structure emerging in the context of special relativity, is the one defined by Kaniadakis \cite{Kaniadakis05} as follows
%
\begin{align}\label{KanEnt}
S_{\{\kappa\}}(p_i)&=\sum_{i=1}^{\Omega_{\mt{max}}^{\mt{K}}}p_i\ln_{\{\kappa\}}(1/p_i),
\end{align}
%
with the $\kappa$-generalized logarithm
%
\begin{align}\label{KanLog}
\ln_{\{\kappa\}}(x)&:=\frac{x^{\kappa}-x^{-\kappa}}{2\kappa},\qquad\rbr{x>0,\,\,\ln_{\{\kappa\ra0\}}(x)=\ln(x)}.
\end{align}
%
The inverse $\kappa$-generalized logarithmic function has the form
%
\begin{align}\label{KappaExp}
\exp_{\{\kappa\}}(x)&=\sqbr{\kappa\,x+\sqrt{1+\kappa^2x^2}}_+^{\frac{1}{\kappa}}
\qquad\rbr{\exp_{\{\kappa\ra0\}}(x)=\exp(x)},
\end{align}
%
which maximizes, according to our previous discussion, the entropy $S_{\kappa}$.
However, computing the inner structure of $S_{\{\kappa\}}$, we obtain
%
\begin{align}\label{DerivKaniadEnt}
\frac{d}{dx}\sqbr{x\ln_{\kappa}(1/x)}&=f_1(\kappa)\ln_{\kappa}(1/x)-f_2(\kappa;x),\\
f_1(\kappa)&=1-\kappa,\\
f_2(\kappa;x)&=x^{\kappa}.
\end{align}
%
As can be seen  the function $f_2(\kappa;x)$ presents explicit dependence on $x$ and thus the variation of the functional $\mc{I}(x=p_i)$ in Eq.~(\ref{JF1}) for $S_{\mc{Q}}=S_{\kappa}$ can not reproduce PDFs of $\kappa$-exponential type.

\subsection{Schw\"ammle-Tsallis $qq'$-Entropy}
%
The application of the $q$-logarithm (\ref{TsallisLogarithmicFunction}) on $q$-multiplied variables gives an ordinary sum of $q$-logarithms applied on each variable separately.
An interesting mathematical question is whether one can define a deformed logarithm whose application on a $q$-product would give a deformed sum of these logarithms.
This point was explored in Ref.~\cite{SchwammleTsallis} by Schw\"ammle and Tsallis who introduced the two parametric
generalized logarithm
%
\begin{align}\label{ST-log}
\ln^{\mt{ST}}_{q,q'}(x)&:=\ln_{q'}^{\mt{T}}\rbr{\exp\rbr{\ln_{q}^{\mt{T}}(x)}}
\qquad \rbr{\ln^{\mt{ST}}_{q\ra1,q'\ra1}(x)=\ln{x}}
\end{align}
%
with its inverse function
%
\begin{align}\label{ST-exp}
\exp^{\mt{ST}}_{q,q'}(x)&=\exp^{\mt{T}}_q\rbr{\ln\rbr{\exp^{\mt{T}}_{q'}(x)}}
\qquad \rbr{\exp^{\mt{ST}}_{q\ra1,q'\ra1}(x)=\exp(x)}.
\end{align}
%
The respective entropic structure based on Eq.~(\ref{ST-log}) is of the form
%
\begin{align}
S^{\mt{ST}}_{q,q'}(p_i)&=\sum_{i=1}^{\Omega_{\mt{max}}^{\mt{ST}}}p_i\ln^{\mt{ST}}_{q,q'}(1/p_i).
\end{align}
%
The derivative of its inner structure leads to the following result
%
\begin{align}\label{ST-deriv-1}%
\frac{d}{dx}\sqbr{x\ln^{\mt{ST}}_{q,q'}(1/x)}&=f_1(q,q';x)\ln^{\mt{ST}}_{q,q'}(1/x)-f_2(q;x),\\
\label{ST-deriv-2}%
f_{1}(q,q';x)&=1-(1-q')x^{q-1},\\
\label{ST-deriv-3}%
f_2(q;x)&=x^{q-1}.
\end{align}
%
Same as in the case of Kaniadakis entropy (\ref{KanEnt}), from Eqs.~(\ref{ST-deriv-1}) - (\ref{ST-deriv-3}) it becomes obvious that the deformed exponential function (\ref{ST-exp}) cannot be obtained from JF.

\subsection{Borges-Roditi $qq'$-Entropy}
%
Another two-parametric trace-form entropy, $S_{q,q'}^{\mt{BR}}=\sum_{i=1}^{\Omega_{\mt{max}}^{\mt{BR}}}p_i\ln_{q,q'}^{\mt{BR}}(1/p_i)$, is the one proposed by Borges and Roditi in Ref. \cite{BorgesRoditi}, based on the deformed logarithm
%
\begin{align}\label{BR-log}
\ln^{\mt{BR}}_{q,q'}(x)&:=\frac{x^{1-q'}-x^{1-q}}{q-q'}
\qquad \rbr{\ln^{\mt{BR}}_{q\ra1,q'\ra1}(x)=\ln{x}}.
\end{align}
%
$S_{q,q'}^{\mt{BR}}$ is generated by applying the $qq'$-generalized derivative introduced by Chakrabarti and Jagannathan \cite{ChakrabartiJagannathan} on the probability functional $g(\alpha):=\sum_ip_i^{\alpha}$, with respect to $\alpha$, and then setting $\alpha=1$.
Eq. (\ref{BR-log}) is not analytically invertible.
The ordinary derivative of the inner structure of $S_{q,q'}^{\mt{BR}}$ yields the following result
%
\begin{align}\label{BR-deriv-1}%
\frac{d}{dx}\sqbr{x\ln^{\mt{BR}}_{q,q'}(1/x)}&=f_1(q,q')\ln^{\mt{BR}}_{q,q'}(1/x)-f_2(q;x),\\
\label{BR-deriv-2}%
f_{1}(q,q')&=q+q',\\
\label{BR-deriv-3}%
f_2(q,q';x)&=\frac{q'x^{q-1}-qx^{q'-1}}{q'-q}.
\end{align}
%
In contrast to Schw\"ammle-Tsallis case, the function $f_1$ does not depend on the variable $x$.
On the other hand, the existence of the $x$-dependence in $f_2$ implies that the type of the probability distribution computed from JF is not the inverse $\ln_{q,q'}^{\mt{BR}}$-function and thus does not preserve extensivity of the respective entropy.
We notice that for $q'\ra1/q$, $S_{q,q'}^{\mt{BR}}$ tends to the one-parametric quantum group entropy introduced by Abe \cite{Abe97}.
Accordingly, the maximum probability distributions of Abe's definition obtained from JF do not optimize quantum group entropy either.

\subsection{Anteneodo-Plastino $\eta$-Entropy}
%
In Ref.~\cite{AnteneodoPlastino} Anteneodo and Plastino have constructed a generalized entropy $S_{\eta}$ whose maximization according to JF yields PDFs of stretched exponential type.
In terms of a deformed logarithm this entropy is given as
%
\begin{align}
S_{\eta}(p_i)&=\sum_{i=1}^{\Omega_{\mt{max}}^{\mt{AP}}}p_i\ln_{\eta}(1/p_i)
\qquad\rbr{S_{\eta\ra1}=\BG},
\end{align}
%
where
%
\begin{align}\label{AP-log}
\ln_{\eta}(x)&:=x\,\Gamma\rbr{\frac{\eta+1}{\eta},\ln{x}}-\Gamma\rbr{\frac{\eta+1}{\eta}},
\qquad \rbr{\ln_{\eta\ra1}(x)=\ln{(x)}}.
\end{align}
%
Considering Eq.~(\ref{AP-log}) we observe that the inverse function of the $\eta$-logarithm is not a stretched exponential one.
Thus, a stretched exponential function does not make the entropy $S_{\eta}$ extensive.
Additionally, the $\eta$-logarithm is not analytically invertible.

\subsection{Thurner-Hanel $gg$-Entropy}
%
Thurner and Hanel in Ref. \cite{ThurnerHanel} have estimated the difficulty of obtaining the proper maximum PDFs for an entropic form within JF created by the term $x\frac{d}{dx}\ln_{\mc{Q}}(1/x)$ in the middle hand side of Eq.~(\ref{derivGenFun}).
In order to solve this problem they added one more term in the BGS-entropy definition, whose structure eliminates the
aforementioned term.
Their entropy definition $S_{gg}$ is given as
%
\begin{align}\label{TH-entropy}
S_{gg}(p_i)&:=-\sum_{i=1}^{\Omega_{\mt{max}}^{gg}}\sqbr{p_i\ln_{gg}(p_i)-\int_0^{p_i}dx\,x\,\frac{d\ln_{gg}(x)}{dx}}+c%
=-\sum_{i=1}^{\Omega_{\mt{max}}^{gg}}\int_0^{p_i}dx\,\ln_{gg}(x)+c
\end{align}
%
where $c$ is a constant.
Indeed, in this case the application of JF on $S_{gg}$ leads to the inverse function of $\ln_{gg}(p_i)$,
namely, $\exp_{gg}(p_i)$.
However, the authors did not take into account that the addition of the integral-term may break the extensivity of $S_{gg}$.
Let us see this situation closer.
For equal probabilities $S_{gg}$ tends to
%
\begin{align}
S_{gg}(\mt{max})&=-\Omega_{\mt{max}}^{gg}\int_0^{1/\Omega_{\mt{max}}^{gg}}dx \ln_{gg}(x)+c.
\end{align}
%
Since $\exp_{gg}$ is the inverse function of $\ln_{gg}$, the maximum configuration function has the form
$\Omega_{\mt{max}}^{gg}=\exp_{gg}\rbr{N\ln_{gg}(W)}$.
Assuming, that $\ln_{gg}(x)$ is integrable with $\int_0^{1/y} dx\ln_{gg}(x)=\tilde{\ln}_{gg}(1/y)$ and $\tilde{\ln}_{gg}(0)=0$, and substituting $Y=N\ln_{gg}(W)$, then we obtain
%
\begin{align}
S_{gg}(\mt{max})&=-\exp_{gg}(Y)\,\tilde{\ln}_{gg}\rbr{1/\exp_{gg}(Y)}+c.
\end{align}
%
Extensivity is succeeded in $S_{gg}(\mt{max})$ when the following condition is  fulfilled
%
\begin{align}\label{Condition1}
\tilde{\ln}_{gg}\rbr{1/\exp_{gg}(Y)}&=-Y/\exp_{gg}(Y)\quad\Longrightarrow\quad \frac{d\exp_{gg}(Y)}{dY}=
\frac{\exp_{gg}(Y)}{\ln_{gg}(1/\exp_{gg}(Y))+Y}.
\end{align}
%
Under specific assumptions the authors presented analytical expressions of the functions $\exp_{gg}$ and $\ln_{gg}$ given as follows
%
\begin{align}\label{TH-exp}%
\exp_{gg}(\gamma;x)&:=\exp\sqbr{-(\pi\gamma^2)^{-1}\sqbr{\mt{erf}^{-1}(2\gamma\sqrt{-x})}^2}
\qquad \rbr{\exp_{gg}(\gamma\ra0;x)=\exp(x)},\\
\label{TH-log}%
\ln_{gg}(\gamma;x)&:=-\sqbr{(2\gamma)^{-1}\mt{erf}\rbr{\gamma\sqrt{-\pi\ln(x)}}}^2
\qquad \rbr{\ln_{gg}(\gamma\ra0;x)=\ln(x)}.
\end{align}
%
One can easily verify that the generalized definitions (\ref{TH-exp}) and (\ref{TH-log}) do not satisfy the condition
(\ref{Condition1}) and thus $S_{gg}(\mt{max})$ depending on $\ln_{gg}(\gamma;x)$ does not become extensive for PDFs of $\exp_{gg}(\gamma;x)$ type. We would like to stress that the structure of entropy (\ref{TH-entropy}) has been first introduced in Ref. \cite{Abe03JPA} based on a different context.

\section{Conclusions}
%
We have shown that the application of Jaynes' Formalism on an arbitrary generalized entropic structure may yield incorrect types of maximum entropy probability distributions.
There are two necessary conditions that must be fulfilled, in order to obtain proper results from the aforementioned
formalism.
The first condition is related to the structure of a generalized entropy definition $S_{\mc{Q}}$ depending on a parameter set $\mc{Q}=\{Q_i\}_{i=1,\ldots,n}$, which must be of trace-form, $S_{\mc{Q}}=\sum_i\Lambda_{\mc{Q}}(p_i)=\sum_ip_i(\Lambda_{\mc{Q}}(p_i)/p_i)=\sum_ip_i\ln_{\mc{Q}}(1/p_i)$ with
$\ln_{\mc{Q}}(p_i):=p_i\Lambda_{\mc{Q}}(1/p_i)$.
The second condition, given in Eq.~(\ref{derivGenFun}), preserves the very important property of extensivity of the entropy $S_{\mc{Q}}$.
From Eq. (\ref{derivGenFun}) we could determine the explicit structure of the $\mc{Q}$-logarithm $\ln_{\mc{Q}}$, presented in Eq.~(\ref{GenLog}).
This is the only generalized logarithm under which the ordinary extremum constraints of the Boltzmann-Gibbs-Shannon entropy do not change.
For $\mc{Q}=Q_1=q$ and $f_1(\mc{Q})=q$ the deformed logarithm $\ln_{\mc{Q}}$ tends to the one defined by Tsallis and
coworkers and the respective deformed entropy $S_{\mc{Q}}$ is Tsallis entropy $\Ts$.

Inverting the above statement,  we see that the definition of the extremum constraints depends strongly on the structural choice of the generalized entropy.
This implies that the generalization procedure of the Boltzmann-Gibbs-Shannon statistics is not a pure mathematical
concept but it carries physical information.
The physics of a statistical ensemble is projected on the extremum constraints.
Thus, one should be very careful about the choice of the $S_{\mc{Q}}$, since changes in the extremum constraints indicates changes in the physics of the system under consideration.
It becomes evident at this point why the choice of Tsallis entropy as a possible generalization of the Boltzmann-Gibbs
entropy in statistical thermodynamics is so special.

In the frame of Tsallis generalized thermostatistics, based on the entropic structures $\Ts$ and $S_{2-q}^{\mt{T}}$ for the respective parameter ranges $[0,1]$ and $[1,2]$, the correct maximum entropy probability functions ($p_i^{\mt{max}}=1/\exp_q(x)=\exp_{2-q}(-x)$ and $p_i^{\mt{max}}=1/\exp_{2-q}(x)=\exp_{q}(-x)$, respectively),  within Jaynes' Formalism are associated to the \emph{ordinary mean value} definition.
Furthermore, in a recent study \cite{Abe08} Abe showed that the escort probability distributions $g(p_i)=p_i^q/\sum_kp_k^q$, which have been widely used in the literature, are not stable for specific probability perturbations.
Considering the above results we conclude that the introduction of the aforementioned escort probability distributions $g(p_i)$ has no physical hypothesis within generalized thermostatistics.

\section*{Acknowledgments}
We would like to thank E. P. Borges for his very fruitful comments.
This work has been supported by TUBITAK (Turkish Agency) under the Research Project number 108T013.


\end{document}